\documentclass{iaus}
\usepackage{graphicx}

\title[ARCONS] 
{ARCONS: a highly multiplexed superconducting UV to near-IR camera}

\author[Kieran O'Brien]   
{Kieran O'Brien$^1$, Ben Mazin$^1$, Sean McHugh$^1$, \\ Seth Meeker$^1$
 \and Bruce Bumble$^2$}

\affiliation{$^1$Department of Physics, University of California Santa Barbara \\Santa Barbara, California 93106, USA \\ email: {\tt kobrien@physics.ucsb.edu} \\[\affilskip]
$^2$Jet Propulsion Laboratory \\Pasadena, California 91107, USA}

\pubyear{2011}
\volume{285}  
\pagerange{1-1}
\jname{New Horizons in Time Domain Astronomy}
\editors{R.E.M. Griffin, R.J. Hanisch \& R. Seaman, eds.}
\begin{document}

\maketitle

\begin{abstract}
ARCONS, the Array Camera for Optical to Near-infrared Spectrophotometry, was recently commissioned at the Coud\'{e} focus of the 200-inch Hale Telescope at the Palomar Observatory. At the heart of this unique instrument is a 1024-pixel Microwave Kinetic Inductance Detector (MKID), exploiting the Kinetic Inductance effect to measure the energy of the incoming photon to better than several percent. The ground-breaking instrument is lens-coupled with a pixel scale of 0.23$^{\prime\prime}$/pixel, with each pixel recording the arrival time ($<2\,\mu$ sec) and energy of a photon ($\sim 10\%$) in the optical to near-IR (0.4-1.1 microns) range. The scientific objectives of the instrument include the rapid follow-up and classification of the transient phenomena. 
\end{abstract}

\firstsection 
\section{Background}
{\bf The Kinetic Inductance detector:}

\begin{figure}[htbp]
\begin{center}
\scalebox{0.35}{\includegraphics{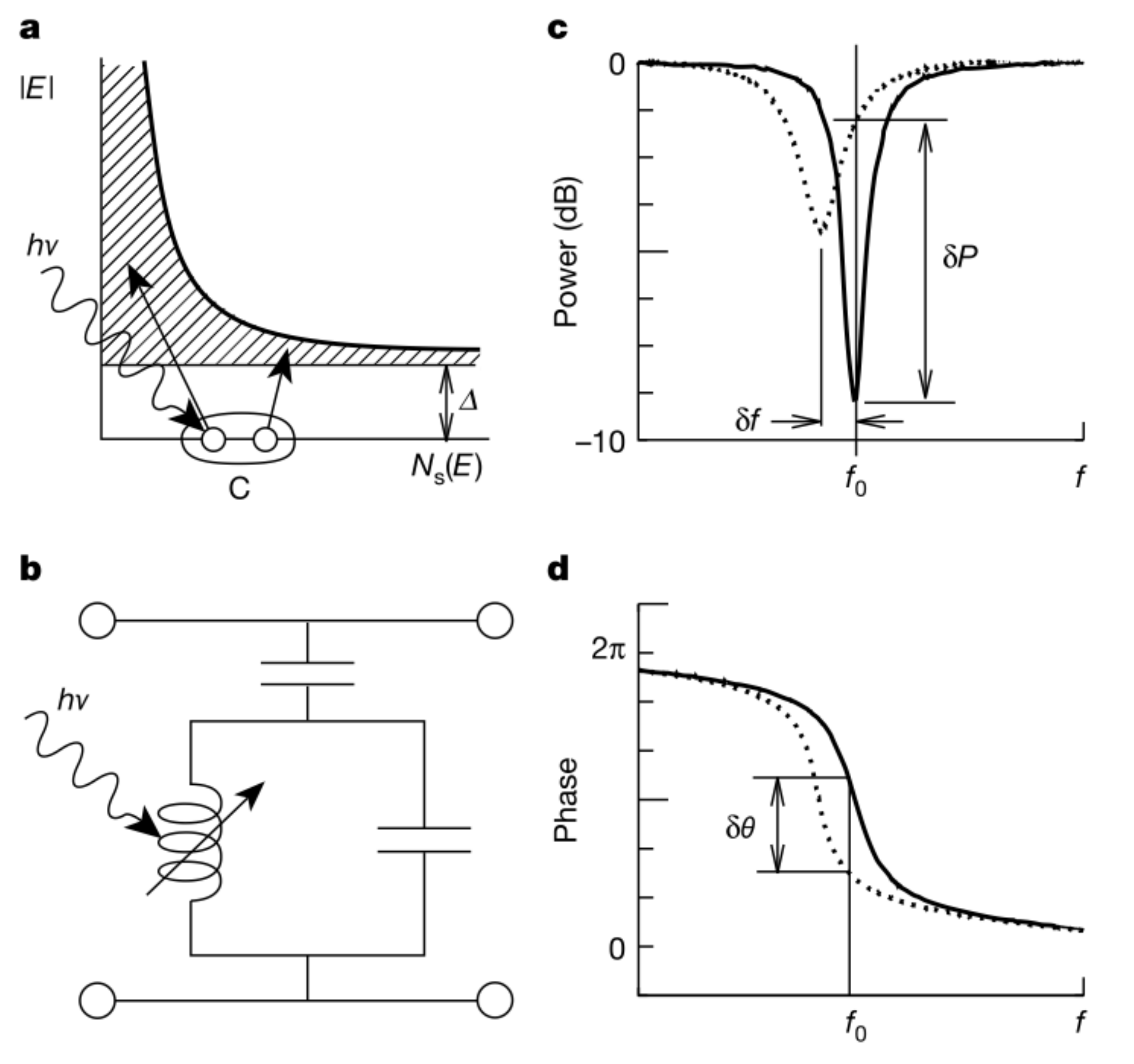}}
\caption{An illustration of the detection principle, from \cite{day2003}. }
\label{kieffect}
\end{center}
\end{figure}

The working principle of the Kinetic Inductance Detector was described in detail in \cite{day2003} and summarized here. Photons with energy $h\nu$ are absorbed in a superconducting film, producing a number of excitations, called `quasiparticles' (Fig~\ref{kieffect}(a)). To sensitively measure these quasiparticles, the film is placed in a high frequency planar resonant circuit (Fig~\ref{kieffect}(b)). Figs~\ref{kieffect}(c,d) show the effect on the amplitude and phase respectively of a microwave excitation signal sent through the resonator. The change in the surface impedance of 
the film following a photon absorption event pushes the resonance to lower frequency and changes its amplitude. If the detector (resonator) is excited with a constant on-resonance microwave signal, we can measure the degree of phase and/or amplitude shift caused by a single incident optical photon.

{\bf Energy resolution:}

As the energy of the incoming photon is many times that necessary to generate a quasiparticle, several thousand quasiparticles are generated by each photon (in contrast to a semiconductor, where the incident photon has an energy only slightly above the band-gap). The degree of phase-shift is related to the number of quasiparticles and hence, the energy of the incoming photon. The maximum theoretical resolution, $R\,(= E/{\delta}E)$ is given by,  
\begin{equation}
R = \frac{1}{2.355}\sqrt{\frac{\eta h \nu}{F \Delta}}
\end{equation}
where $\eta$ is an efficiency factor, F the Fano factor and $\Delta$ the superconducting energy gap.

{\bf Multiplexing scheme:}

By engineering each resonator to have a slightly different resonant frequency, a large (few thousand) number of resonators (pixels) can be simultaneously probed by a comb of frequencies sent down a single coaxial line. In the current scheme we use two coaxial cables with 512 resonators on each line in the range 4-5GHz. 

\section{The Instrument}
 
\begin{figure}[htbp]
\begin{center}
\scalebox{0.9}{\includegraphics{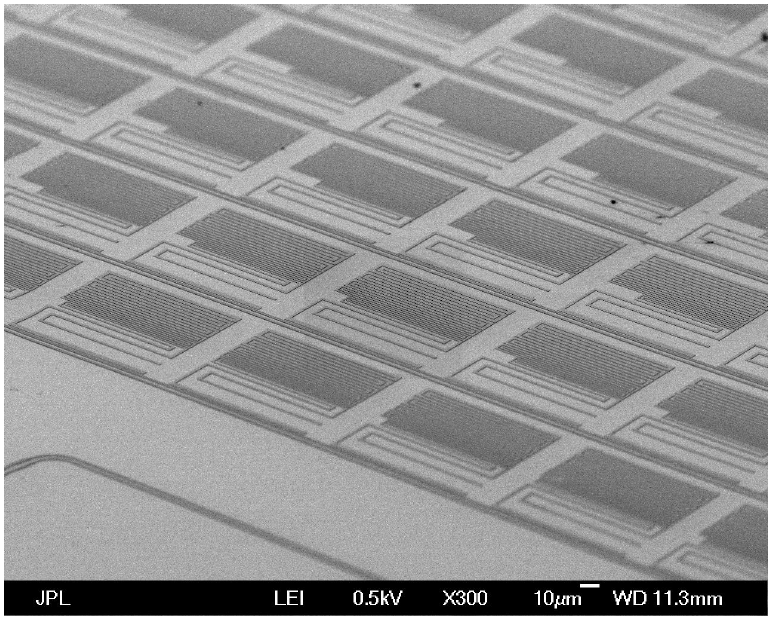}} \scalebox{0.7}{\includegraphics{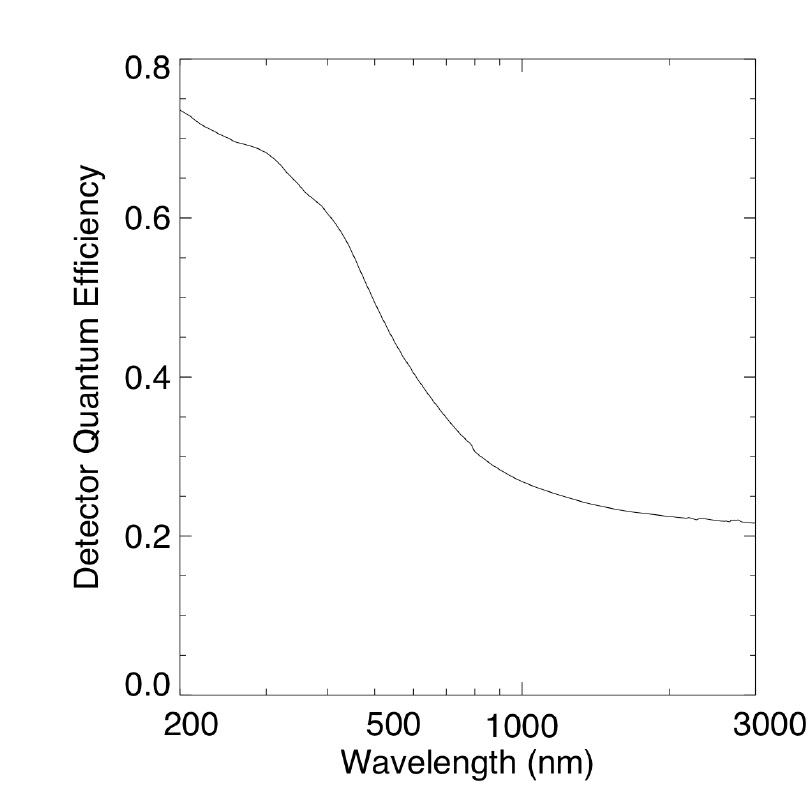}}
\caption{Left, Image of a section of the science array. The individual pixels can been seen to have a slightly different length meandered section in order to tune the resonant frequency, enabling the highly multiplexed read-out. Right, the measured quantum efficiency of the TiN lumped element detector.}
\label{fig2}
\end{center}
\end{figure}

ARCONS (\cite{mazin2010}) uses a cryogen-free ADR to cool an array of Titanium-Nitride (TiN) lumped element MKIDs to a temperature of 85mK (well below the superconductor $T_c$ of $\sim$800mK). The instrument has a hold time of $\sim$12hrs before it needs to be regenerated, which takes $\sim$2hrs. The 32x32 array (shown in Fig.~\ref{fig2}) is on a $100\mu$m pitch, which is behind a $200\mu$m focal length micro-lens array, increasing the fill factor to 64\% and increases the uniformity by concentrating the light on a small region of the inductor. While the MKIDs are sensitive from the UV to mid-IR (see Fig.~\ref{fig2}), the choice of glass cuts them off below 400nm and a filter (ASAHI `super-coldÕ) sets the red limit at 1100nm, limiting the total count-rate (dominated by the sky). The pixels are 0.23$^{\prime\prime}$ on the sky, giving a field-of-view of $\sim$7.5x7.5$^{\prime\prime}$. Each pixel has a R$\sim$12 at 400nm, but was strongly affected by Ôsubstrate eventsÕ where the photon is absorbed in the Si substrate, leading to a breakdown in the relationship between photon energy and phase shift.
 
The pixels are read out using a custom-built software defined radio (SDR) system (\cite{mchugh2011}). In this system a frequency comb is created in software, and up-converted to the required frequency range. This signal is sent along the coax and passes through the MKIDs. The transmitted signal is amplified and then down-converted and digitized via onboard A/D converters. The signal is then `channelizedÕ and the pulse heights (a direct measure of the photon energy) and pulse start times (photon arrival times) are measured in a powerful FPGA.

\section{First Light}
ARCONS was successfully commissioned during 4 nights in July/August 2011.  The through-put was as expected. We observed a broad range of `science demonstrationÕ targets, including interacting binaries (AM Cvns, LMXBs, short period eclipsing sources), QSOs (for low resolution redshift measurements), supernovae (Type Ia and Type II) and the Crab pulsar. 

\begin{figure}[htbp]
\begin{center}
\scalebox{0.375}{\includegraphics{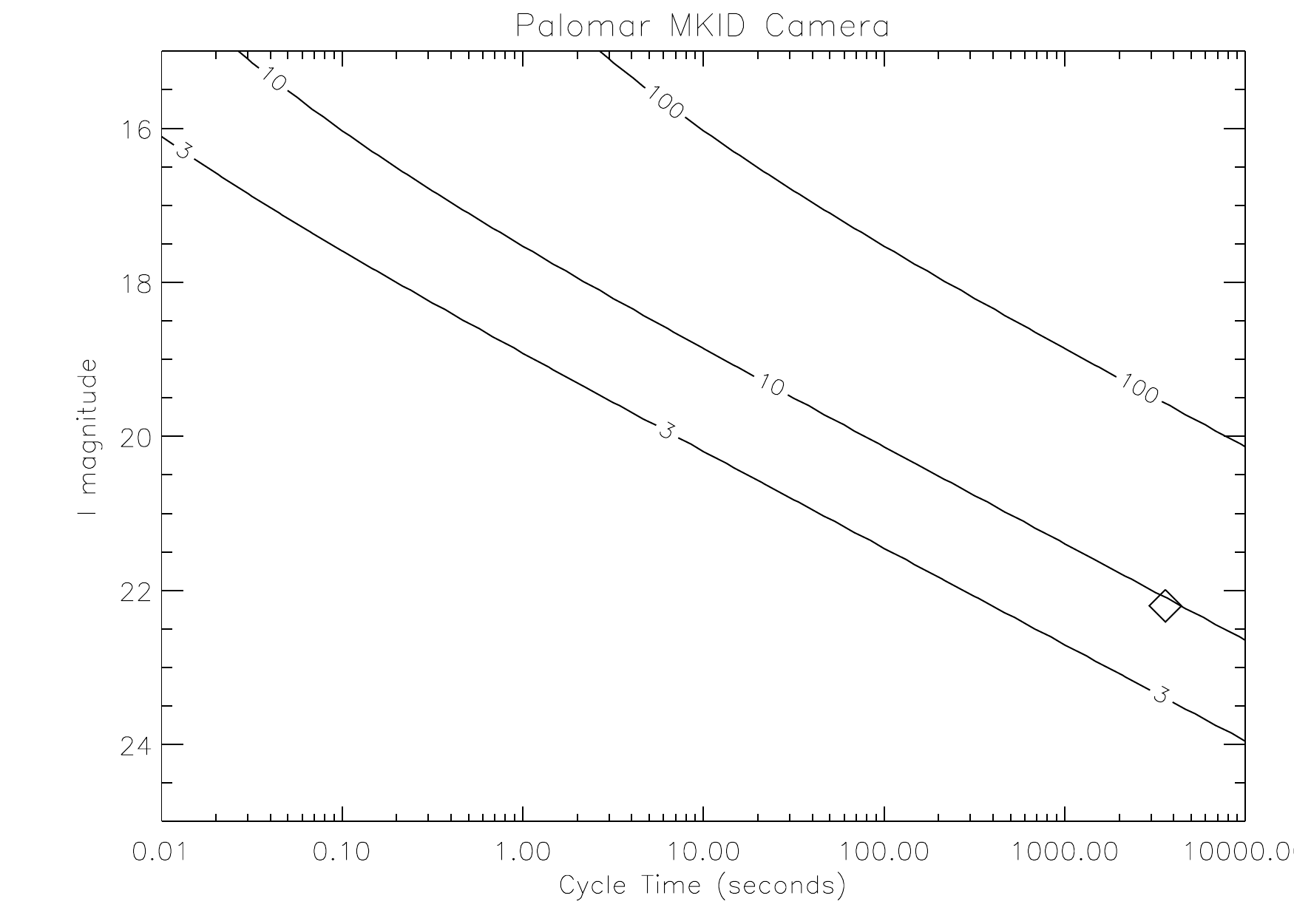}} \scalebox{0.33}{\includegraphics{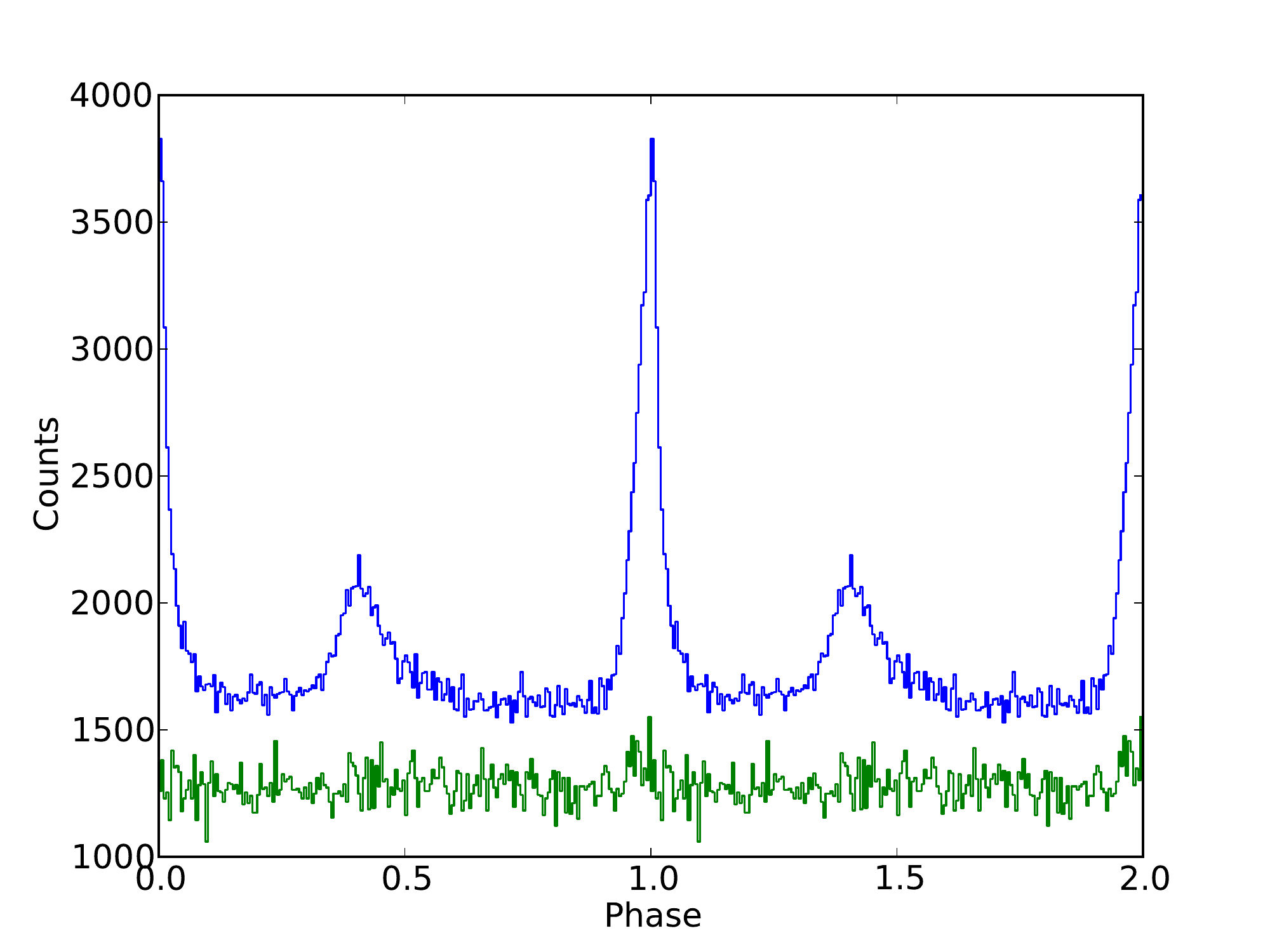}}
\caption{Left, predicted sensitivity of ARCONS at Palomar with contours for signal-to-noise ratios of 3, 10 and 100. The diamond shows a measurement from the 2011 commissioning run. Right, the phase-folded Crab Pulsar lightcurve (repeated for 2 cycles of the 33 msec spin period for clarity) from a 20\,sec twilight observation. The sky variability is also shown (offset).  }
\label{fig3}
\end{center}
\end{figure}

\section{Science Goals}

\begin{enumerate}
\item {\bf Rapid characterization of transients:} ARCONS will allow us to determine the spectrum of a transient source as it evolves. The small field, read-noise free IFU will allow point-and-shoot deep spectroscopic observations. This will enable better marshaling of follow-up on large aperture telescopes.
\item {\bf Time resolved Opt/IR observations of pulsars:} Detailed time-resolved spectra, especially if combined with simultaneous radio (GBT) and gamma-ray (Fermi) observations should reveal details about pulsar emission and the structure of the magnetosphere that will allow us to differentiate between models of optical emission. 
\item {\bf Characterization of short-period variables:} There are a growing number of short period variables being discovered by sky surveys and this number will increase with future large surveys, most notably LSST. These discoveries require detailed follow-up observations in order to characterize the components of the binary. ARCONS is ideally suited to such observations, offering zero dead-time, low-resolution spectroscopy. 
\item {\bf Redshift determination via spectro-imaging:} We will use the intrinsic energy resolution of the MKIDs to determine the redshift of galaxies out to a redshift $\sim$4, with a high degree of accuracy. A spectral resolution, R, of 20 in the UV with ARCONS translates into 14 wavelength resolution elements in the wavelength range of the camera. The advantage of ARCONS over multi-filter photometry, such as that employed by COMBO-17, is that all wavelengths are observed simultaneously (thus reducing the exposure time considerably) and through the same conditions (seeing, sky transparency), making the analysis less complicated. 
\end{enumerate}
 
\section{The Future}
With the information and experience we have gained from the recent commissioning run, we have identified a number of upgrades that combined will improve the instrument significantly and take it from a demonstration instrument to a front-line scientific instrument. These include increasing the pixel-scale to better match the median seeing at Palomar and give us a wider field of view, increasing the number of pixels and bandwidth of the read-out electronics to read-out a 2048 pixel array and improving our calibration scheme. In the longer-term, we are developing an anti-reflection scheme to improve the QE of the detectors. 
MKIDs promise to revolutionize many areas of astronomy, not limited to the time domain, as they are capable of producing large area, read-noise free integral-field spectroscopy without the need for complicated optical systems. In the time domain, they will offer deep, low resolution integral field spectroscopy with the advantage of time-tagging the arrival time of each photon.

\section{Acknowledgements}
We would like to thank the management and staff of the Palomar Observatory for their hard work and support during the commissioning of ARCONS. This material is based upon work supported by the National Aeronautics and Space Administration under Grant NNX09AD54G, issued through the Science Mission Directorate, Jet Propulsion LabÕs Research \& Technology Development Program, and a grant from the W.M. Keck Institute for Space Studies. Part of the research was carried out at the Jet Propulsion Laboratory, California Institute of Technology, under a contract with the National Aeronautics and Space Administration.

\end{document}